\documentclass[11pt,a4paper]{article}
\usepackage{amsmath}
\usepackage{amssymb}
\usepackage{amsthm}
\usepackage{epsfig}
\usepackage{times}
\usepackage{euler}
\usepackage{wrapfig}
\usepackage{epsfig}
\usepackage{a4}
\usepackage[hang,stable]{footmisc}
\begin{document}
\ifx\href\undefined\else\hypersetup{linktocpage=true}\fi

\newtheorem{theorem}{Theorem}
\newtheorem{proposition}{Proposition}
\theoremstyle{definition}
\newtheorem{Requirement}{Requirement}
\newtheorem{Remark}{Remark}

\title{
Electron Spin\\ 
or           \\ 
``Classically Non-Describable Two-Valuedness''}
\author{
Domenico Giulini            \\
Max-Planck-Institute for Gravitational Physics  \\
Albert-Einstein-Institute                       \\
Am M\"uhlenberg 1                               \\
D-14476 Golm/Potsdam}

\date{}

\maketitle

\begin{abstract}
\noindent
In December 1924 Wolfgang Pauli proposed the idea of an inner degree 
of freedom of the electron, which he insisted should be thought of
as genuinely quantum mechanical in nature. Shortly thereafter Ralph
Kronig and, independently, Samuel Goudsmit and George Uhlenbeck 
took up a less radical stance by suggesting that this degree of 
freedom somehow corresponded to an inner rotational motion, though 
it was unclear from the very beginning how literal one was actually 
supposed to take this picture, since it was immediately recognised 
(already by Goudsmit and Uhlenbeck) that it would very likely lead 
to serious problems with Special Relativity if the model were to 
reproduce the electron's values for mass, charge, angular 
momentum, and magnetic moment. However, probably due to the then
overwhelming impression that classical concepts were generally 
insufficient for the proper description of microscopic phenomena,    
a more detailed reasoning was never given. In this contribution I 
shall investigate in some detail what the restrictions on the 
physical quantities just mentioned are, if they are to be 
reproduced by rather simple classical models of the electron 
within the framework of Special Relativity. It turns out that 
surface stresses play a decisive role and that the question of 
whether a classical model for the electron does indeed contradict 
Special Relativity can only be answered on the basis of an 
\emph{exact} solution, which has hitherto not been given.  
\end{abstract}

\newpage
\begin{small}
\setcounter{tocdepth}{3}
\tableofcontents
\end{small}

\newpage

\section{Introduction}
\label{sec:Intro}
The discovery of electron spin is one of the most interesting 
stories in the history of Quantum Mechanics; told e.g. 
in van der Waerden's contribution to the Pauli Memorial Volume 
(\cite{Fierz.Weisskopf:PauliMemorial}, pp.\,199-244), in 
Tomonaga's book \cite{Tomonaga:StoryOfSpin}, and also in various 
first-hand reports \cite{Uhlenbeck:1976}\cite{Goudsmit:1976}
\cite{Pais:1989}. This story also bears fascinating relations to 
the history of understanding Special Relativity. One such relation 
is given by Thomas' discovery of what we now call ``Thomas precession'' 
\cite{Thomas:1926}\cite{Thomas:1927}, which explained for the first 
time the correct magnitude of spin-orbit coupling and hence the 
correct magnitude of the fine-structure split of spectral lines,
and whose mathematical origin can be traced to precisely 
that point which marks the central difference between the Galilei 
and the Lorentz group (this is e.g. explained in detail in 
Sects.\,4.3-4.6 of \cite{Giulini:2006b}). In the present paper  
I will dwell a little on another such connection to Special 
Relativity. 

As is widely appreciated, Wolfgang Pauli is a central figure, perhaps
\emph{the} most central figure, in the story of spin . Being the inventor 
of the idea of an inner (quantum mechanical) degree of freedom of the 
electron, he was at the same time the strongest opponent to attempts 
to relate it to any kind of interpretation in terms of 
kinematical concepts that derive from the picture of an extended 
material object in a state of rotation. To my knowledge, Pauli's 
hypothesis of this new intrinsic feature of the electron, which he 
cautiously called ``a classical non-describable two valuedness'', was 
the first instance where a quantum-mechanical degree of 
freedom was claimed to exist without a corresponding classical one. 
This seems to be an early attempt to walk without 
the crutches of some `correspondence principle'. Even though the 
ensuing developments seem to have re-installed -- mentally at least -- 
the more classical notion of a spinning electron through the ideas of 
Ralph Kronig (compare section~4 of van\,der\,Waerden's contribution 
to \cite{Fierz.Weisskopf:PauliMemorial}, pp.\,209-216) 
and, independently,  Samuel Goudsmit and George Uhlenbeck 
\cite{Goudsmit.Uhlenbeck:1925}\cite{Goudsmit.Uhlenbeck:1926}, Pauli 
was never convinced, despite the fact that he lost the battle against 
Thomas\footnote{At this point Frenkel's remarkable contribution 
\cite{Frenkel:1926} should also be mentioned, which definitely improves 
on Thomas' presentation and which was motivated by Pauli sending Frenkel 
Thomas' manuscript, as Frenkel acknowledges in footnote\,1 on p.\,244 
of \cite{Frenkel:1926}. A more modern account of Frenkel's work is 
given in \cite{Ternov.Bordovitsyn:1980}.} and declared ``total surrender'' 
in a letter to Bohr written on March 12. 1926 (\cite{Pauli:SC}, 
Vol.\,I, Doc.\,127, pp.\,310). For Pauli the spin of the electron 
remained an abstract property which receives its ultimate and irreducible 
explanation in terms of group theory, as applied to the subgroup\footnote{It 
is more correct to speak of the conjugacy class of subgroups of spatial 
rotations, since there is no (and cannot be) a single distinguished 
subgroup group of `spatial' rotations in Special Relativity.} of 
spatial rotations (or its double cover) within the full symmetry 
group of space-time, may it be the Galilei or the Lorentz group 
(or their double cover).\footnote{Half-integer spin representations only 
arise either as proper ray-representations (sometimes called `double-valued' 
representations) of spatial rotations $SO(3)$ or as faithful true 
representations (i.e. `single-valued') of its double-cover group 
$SU(2)$, which are subgroups of the Galilei and Lorentz groups or 
their double-cover groups respectively.} In this respect, Pauli's 
1946 Nobel Lecture contains the following instructive passage (here 
and throughout this paper I enclose my annotations to quotes within 
square brackets): 
\begin{quote}
Although at first I strongly doubted the correctness of this idea [of the 
electron spin in the sense of Kronig, Goudsmit and Uhlenbeck] because of 
its classical-mechanical character, I was finally converted to it by 
Thomas' calculations on the magnitude of doublet splitting. On the other 
hand, my earlier doubts as well as the cautions expression $\ll$classically 
non-describable two-valuedness$\gg$ experienced a certain verification during 
later developments, since Bohr was able to show on the basis of wave 
mechanics that the electron spin cannot be measured by classically 
describable experiments (as, for instance, deflection of molecular 
beams in external electromagnetic fields) and must therefore be 
considered as an essential quantum-mechanical property of the 
electron.\footnote{At this point Pauli refers to the reports of the 
Sixth Physics Solvay Conference 1932.}
(\cite{Pauli:NobelLecture}, p.\,30)            
\end{quote}

\begin{wrapfigure}{r}{0.6\linewidth}
\vspace{-0.3cm}
\centering\epsfig{figure=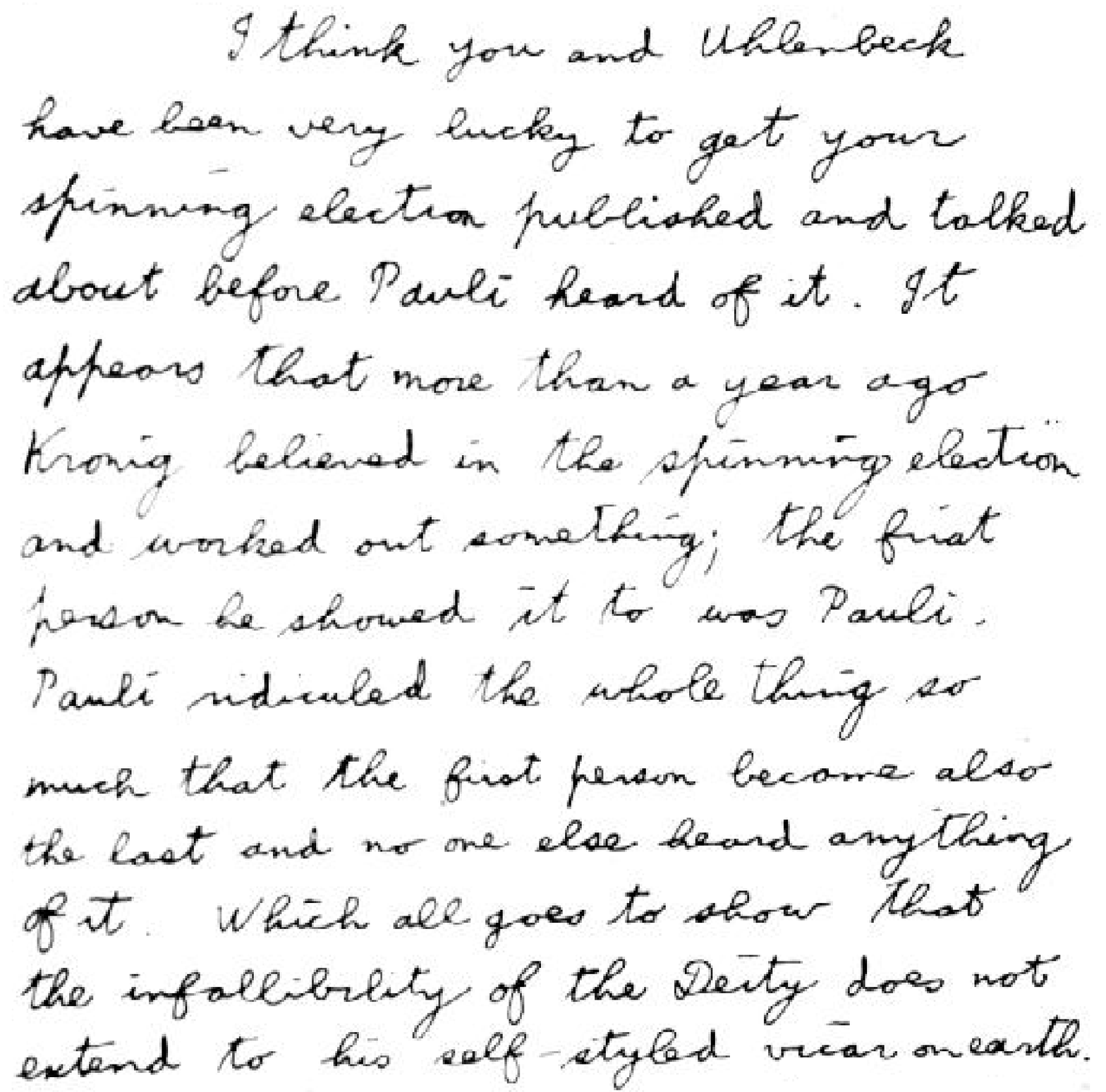, width=\linewidth}
\caption{\label{fig:ThomasLetter}\footnotesize
Part of a letter by L.H.~Thomas to S.~Goudsmit dated March 25th 1926,
taken from \cite{Goudsmit:1971-Lecture}}
\end{wrapfigure}
This should clearly not be misunderstood as saying that under the impression 
of Thomas' calculations Pauli accepted spin in its `classical-mechanical'  
interpretation. In fact, he kept on arguing fiercely against what in a letter 
to Sommerfeld from December 1924 he called ``model prejudices'' (\cite{Pauli:SC}, 
Vol.\,I, Doc.\,72, p.\,182) and did not refrain from ridiculing the upcoming 
idea of spin from the very first moment (cf. Fig.\,1). What Pauli accepted 
was the idea of the electron possessing an intrinsic magnetic moment and 
angular momentum, 
the latter being interpreted \emph{exclusively} in a formal fashion through 
its connection with the generators of the subgroup of rotations within the 
Lorentz group, much like we nowadays view it in modern relativistic 
field theory. To some extent it seems fair to say that, in this case, 
Pauli was a pioneer of the modern view according to which abstract 
concepts based on symmetry-principles are seen as primary, whereas their 
concrete interpretation in terms of localised material structures, to 
which e.g. the kinematical concept of `rotation' in the proper sense applies, 
is secondary and sometimes even dispensable. But one should not forget that 
this process of emancipation was already going on in connection with the 
notion of classical fields, as Einstein used to emphasise, e.g., in his 1920 
Leiden address ``Ether and the Theory of 
Relativity''\footnote{German original: \"Ather und Relativit\"atstheorie.} 
(\cite{Einstein:CP}, Vol.\,7, Doc.\,38, pp.\,308-320). We will come back to this 
point below.\footnote{The case of a classical electromagnetic field is 
of particular interesting insofar as the suggestive picture provided by 
Faraday's lines of force, which is undoubtedly helpful in many cases, 
also provokes to view these lines as objects in space, like ropes 
under tension, which can be attributed a variable state of motion. 
But this turns out to be a fatal misconception.}

Besides being sceptical in general, Pauli once also made a \emph{specific} 
remark as to the inadequateness of classical electron models; that was 
three years after Thomas' note, in a footnote in the addendum to his 
survey article ``General Foundations of the Quantum Theory of Atomic 
Structure''\footnote{German original: Allgemeine Grundlagen der 
Quantentheorie des Atombaues.}, that appeared 1929 as chapter 29 in 
`M\"uller-Pouillets Lehrbuch der Physik'. There he said:
\begin{quote}
Emphasising the kinematical aspects one also speaks of the `rotating electron'
(English `spin-electron'). However, we do not regard the conception of a 
rotating material structure to be essential, and it does not even recommend 
itself for reasons of superluminal velocities one then has to accept.
(\cite{Pauli:CSP}, Vol.\,1, pp.\,721-722, footnote\,2) 
\end{quote}  
Interestingly, this is precisely the objection that, according to Goudsmit's 
recollections~\cite{Goudsmit:1971-Lecture}, Lorentz put forward when 
presented with Goudsmit's and Uhlenbeck's idea by Uhlenbeck, and which 
impressed Uhlenbeck so much that he asked Ehrenfest for help in withdrawing 
the already submitted paper~\cite{Goudsmit:1971-Lecture}. He did not 
succeed, but the printed version contains at least a footnote pointing 
out that difficulty:
\begin{quote}
The electron must now assume the property (a) [a $g$-factor of 2],
which \textsc{Land\'e}  attributed to the atom's core, and which is hitherto
not understood. The quantitative details may well depend on the choice 
of model for the electron. [...] Note that upon quantisation of that 
rotational motion [of the spherical hollow electron], the equatorial 
velocity will greatly exceed the velocity of light.
(\cite{Goudsmit.Uhlenbeck:1925}, p.\,954)
\end{quote} 
This clearly says that a classical electron model cannot reproduce 
the observable quantities, mass, charge, angular momentum, 
and magnetic moment, without running into severe contradictions 
with Special Relativity.\footnote{The phrase ``upon quantisation'' 
in the above quotation is to be understood quantitatively, i.e. 
as ``upon requiring the spin angular-momentum to be of magnitude 
$\hbar/2$ and the magnetic moment to be one magneton ($g=2$)''.}
The electron model they had in mind was that developed by Abraham 
in his 1903 classic paper on the ``Principles of Electron 
Dynamics'' \cite{Abraham:1902} (cited in footnote\,2 on p.\,954 of 
\cite{Goudsmit.Uhlenbeck:1925}). Since then it has become 
standard textbook wisdom that classical electron models necessarily 
suffer from such defects (compare, e.g., \cite{Born:AtomicPhysics}, 
p.\,155) and that, even in quantum mechanics, ``the idea of the rotating 
electron is not be taken literally'', as Max Born once put it 
(\cite{Born:AtomicPhysics}, p.\,188). Modern references iterate 
this almost verbatim:
\begin{quote}
The term `electron spin' is not to be taken literally in the classical 
sense as a description of the origin of the magnetic moment described above. 
To be sure, a spinning sphere of charge can produce a magnetic moment, but 
the magnitude of the magnetic moment obtained above cannot be reasonably 
modelled by considering the electron as a spinning sphere.\\ 
(Taken from
$\langle$ http://hyperphysics.phy-astr.gsu.edu/hbase/spin.html$\rangle$)
\end{quote}

In this contribution I wish to scrutinise the last statement. 
This is not done in an attempt to regain respect for classical electron 
models for modern physics, but rather to illuminate in some detail a 
specific and interesting case of the (well know) general fact that 
progress is often driven by a strange mixture of good and bad arguments, 
which hardly anybody cares to separate once progress is seen to 
advance in the `right direction'. Also, the issues connected with 
an inner rotational motion of the electron are hardly mentioned 
in the otherwise very detailed discussion of classical electron 
theories in the history-of-physics literature (compare 
\cite{Miller:1973}\cite{Janssen.Mecklenburg}). Last but not least, 
the present investigation once more emphasises the importance of 
special-relativistic effects due to stresses, which are not 
necessarily connected with large velocities, at least in a 
phenomenological description of matter. But before giving a 
self-contained account, I wish to recall Pauli's classic paper 
of December 1924, where he introduced his famous ``classically 
non-describable two-valuedness''.

\section{A classically non-describable two-valuedness}
\subsection{Preliminaries}
\label{sec:Preliminaries}
We begin by recalling the notion of \emph{gyromagnetic ratio}. Consider a 
(not necessarily continuous) distribution of mass and charge in the 
context of pre-Special-Relativistic physics, like, e.g., a charged fluid 
or a finite number of point particles. Let $\vec v(\vec x)$ denote 
the corresponding velocity field with respect to an inertial frame 
and $\rho_q$ and $\rho_m$ the density distributions of electric 
charge and mass corresponding to the total charge $q$ 
and mass $m_0$ respectively. The total angular momentum is given by 
($\times$ denotes the antisymmetric vector product)
\begin{equation}
\label{eq:IntroAngMom}
\vec J=\int d^3x\,\rho_m(\vec x)\bigl(\vec x\times\vec v(\vec x)\bigr)\,.
\end{equation}
The electric current distribution, $\vec j(\vec x):=\rho_q\vec v(\vec x)$, is the 
source of a magnetic field which at large distances can be approximated by 
a sum of multipole components of increasingly rapid fall-off for large 
distances from the source.
The lowest possible such component is the dipole. It has the slowest 
fall-off (namely $1/r^3$) and is therefore the dominant one at large 
distances.  (A monopole contribution is absent due to the lack of magnetic 
charges.) The dipole field is given by\footnote{We use 
SI units throughout so that the electric and magnetic constants 
$\varepsilon_0$ and $\mu_0$ will appear explicitly. Note that 
$\varepsilon_0\mu_0=1/c^2$ and that 
$\mu_0=4\pi\cdot 10^{-7}\,\mathrm{kg\cdot m\cdot C^{-2}}$
\emph{exactly}, where C stands for `Coulomb', the unit of charge.}
\begin{equation}
\label{eq:MagnDipole}
\vec B_{dipole}(\vec x)
:=\left(\frac{\mu_0}{4\pi}\right)\,
\frac{3\vec n(\vec n\cdot\vec M)-\vec M}{r^3}\,,
\end{equation}
where $r:=\Vert\vec x\Vert$, $\vec n:=\vec x/r$ and where $\vec M$ 
denotes the magnetic dipole moment of the current distribution, which 
is often (we shall follow this) just called \emph{the} magnetic moment:
\begin{equation}
\label{eq:DefMagnMoment}
\vec M:=\tfrac{1}{2}\int d^3x\, \rho_q(\vec x)\,
\bigl(\vec x\times\vec v(\vec x)\bigr)\,.
\end{equation}
Note the similarity in structure to (\ref{eq:IntroAngMom}), except for  
the additional factor of 1/2 in front of~(\ref{eq:DefMagnMoment}).

The \emph{gyromagnetic ratio} of a stationary mass and charge 
current-distribution , $R_g$, is defined to be the ratio of the 
moduli of $\vec M$ and $\vec J$:
\begin{equation}
\label{eq:DefGyrRatio}
R_g:=\frac{\Vert\vec M\Vert}{\Vert\vec J\Vert}\,.
\end{equation}
We further define a dimensionless quantity $g$, called the 
\emph{gyromagnetic factor}, by 
\begin{equation}
\label{eq:DefGyrFactor}
R_g=:g\,\frac{q}{2m_0}\,.
\end{equation}

These notions continue to make sense in non-stationary situations 
if $\vec M$ and $\vec J$ are slowly changing (compared 
to other timescales set by the given problem), or in (quasi) periodic 
situations if $\vec M$ and $\vec J$ are replaced by their time 
averages, or in mixtures of those cases where, e.g., $\vec J$
is slowly changing and $\vec M$ rapidly precesses around $\vec J$
(as in the case discussed below). 

An important special case is given if charge and mass distributions 
are strictly proportional to each other, i.e., 
$\rho_q(\vec x)=\lambda\rho_m(\vec x)$, where $\lambda$ is independent 
of $\vec x$. Then we have 
\begin{equation}
\label{eq:ProptoDensities}
R_g=\frac{q}{2m_0}\Rightarrow g=1\,.
\end{equation}
In particular, this would be the case if charge and mass carriers were 
point particles of the same charge-to-mass ratio, like $N$ particles of 
one sort, where 
\begin{equation}
\label{eq:ProptoDensitiesPointParticles}
\rho_q(\vec x)=\frac{q}{N}\,
 \sum_{i=1}^N\delta^{\scriptscriptstyle(3)}(\vec x-\vec x_i)
\qquad\text{and}\qquad
\rho_m(\vec x)=\frac{m_0}{N}\,
 \sum_{i=1}^N\delta^{\scriptscriptstyle(3)}(\vec x-\vec x_i)\,.
\end{equation}
After these preliminaries we now turn to Pauli's paper. 

\subsection{Pauli's paper of December 1924}
On December 2nd 1924, Pauli submitted a paper entitled 
``On the influence of the velocity dependence of the electron 
mass upon the Zeeman effect''\footnote{German original: \"Uber den 
Einflu\ss\ der Geschwindigkeitsabh\"angigkeit der Elektronenmasse 
auf den Zeemaneffekt.} (\cite{Pauli:CSP}, Vol.\,2, pp.\,201-213) to 
the Zeitschrift f\"ur Physik. In that paper he starts with the general 
observation that for a point particle of rest-mass $m_0$ and 
charge $q$, moving in a bound state within a spherically symmetric 
potential, the velocity dependence of mass, 
\begin{equation}
\label{eq:MassVelDep}
m=m_0/\sqrt{1-\beta^2}\,,
\end{equation}
affects the gyromagnetic ratio. Here $\beta:=v/c$, where $v:=\Vert\vec v\Vert$.
The application he aims for is the 
anomalous Zeeman effect for weak magnetic fields, a topic on which 
he had already written an earlier paper, entitled 'On the Rules 
of the anomalous Zeeman Effect'\footnote{German original: \"Uber 
die Gesetzm\"a\ss igkeiten des anomalen Zeemaneffekts.}
(\cite{Pauli:CSP}, Vol.\,2, pp.\,151-160), in which he pointed out 
certain connections between the weak-field case and the theoretically 
simpler case of a strong magnetic field. Note that ``weak'' and ``strong'' 
here refers to the standard set by the inner magnetic field caused by 
the electrons orbital motion, so that ``weak'' here means that the Zeeman 
split is small compared to the fine-structure.

Since the charge is performing a quasi periodic motion\footnote{Due to 
special-relativistic corrections, the bound orbits of a point charge 
in a Coulomb field are not closed. The leading order perturbation
of the ellipse that one obtains in the Newtonian approximation 
is a prograde precession of its line of apsides.}, its magnetic moment 
due to its orbital motion is given by the time average (I will denote 
the time average of a quantity $X$ by $\langle X\rangle$) 
\begin{equation}
\label{eq:PauliMagnMoment}
\langle\vec M\rangle=q\,\langle\vec x\times\vec v\rangle/2\,.
\end{equation}
On the other hand, its angular momentum is given by 
\begin{equation}
\label{eq:PauliAngMom}
\vec J=m\,(\vec x\times\vec v)
=m_0\,(\vec x\times\vec v)/\textstyle{\sqrt{1-\beta^2}}\,.
\end{equation}
It is constant if no external field is applied and slowly 
precessing around the magnetic field direction if such a 
field is sufficiently weak, thereby keeping a constant 
modulus. Hence we can write    
\begin{equation}
\label{eq:xTimesvAverage}
\langle\vec x\times\vec v\rangle
=\frac{\vec J}{m_0}\left\langle\textstyle{\sqrt{1-\beta^2}}\right\rangle\,,
\end{equation}
where the averaging period is taken to be long compared to the orbital 
period of the charge, but short compared to the precession period of 
$\vec J$ if an external magnetic field is applied. This gives 
\begin{equation}
\label{eq:PauliGyromRatio}
\frac{\Vert\langle\vec M\rangle\Vert}{\Vert\vec J\Vert}
=\frac{\vert q\vert}{2m_0}\,\gamma\,,
\end{equation}
where\footnote{This is Pauli's notation. Do not confuse this $\gamma$ 
with the Lorentz factor $1/\sqrt{1-\beta^2}$, which nowadays is usually 
abbreviated by $\gamma$, though not in the present paper.}  
\begin{equation}
\label{eq:DefGamma}
\gamma:=
\left\langle\textstyle{\sqrt{1-\beta^2}}\right\rangle\,.
\end{equation}

More specifically, Pauli applies this to the case on an electron in 
the Coulomb field of a nucleus. Hence $m_0$ from now on denotes the 
electron mass. Its charge is $q=-e$, and the charge of the nucleus 
is $Ze$. Using the virial theorem, he then gives a very simple 
derivation of 
\begin{equation}
\label{eq:PauliGammaAverage-1}
\gamma\,=\,1+W/m_0c^2\,,
\end{equation}
where $W$ denotes the electron's total energy (kinetic plus potential). 
For the quantised one-electron problem, an explicit expression for 
$W$ in terns of the azimuthal quantum number $k$ ($j+1$ in modern notation,
where $j$ is the quantum number of orbital angular-momentum) 
and the principal quantum number $n$ ($n=n_r+k$, where $n_r$ is the 
radial quantum number) was known since Sommerfeld's 1916 explanation 
of fine structure (see, e.g., \cite{Sommerfeld:1916b}, p.\,53, formula (17)). 
Hence Pauli could further write:
\begin{equation}
\label{eq:PauliGammaAverage-2}
\gamma\,=\,
\left\{1+\frac{\alpha^2 Z^2}{\Bigl(n-k+\sqrt{k^2-\alpha^2Z^2}\,\Bigr)^2}
\right\}^{-1/2}
\ \approx\ 1-\frac{\alpha^2 Z^2}{2n^2}\,,
\end{equation}
where the approximation refers to small values of $\alpha^2Z^2$ and
where $\alpha:=e^2/4\pi\varepsilon_0\hbar c\approx 1/137$ is the 
fine-structure constant. For higher $Z$ one obtains 
significant deviations from the classical value $\gamma=1$. 
For example, $Z=80$ gives $g=0.812$. 

The relativistic correction factor $\gamma$ affects the angular 
frequency\footnote{We will translate all proper frequencies in 
Pauli's paper into angular frequencies. Hence there are 
differences in factors of $2\pi$. This is also related to 
our usage of $\hbar:=h/2\pi$ rather than $h$ (Planck's constant).} 
with which the magnetic moment created by the electron's orbital 
motion will precess in 
a magnetic field of strength $B$. This angular frequency is now 
given by $\gamma\omega_0$, where $\omega_0$ is the Larmor 
(angular) frequency: 
\begin{equation}
\label{eq:DefLarmorFrequency}
\omega_0=g_e\frac{eB}{2m_0}\,.
\end{equation}
Here we explicitly wrote down the gyromagnetic ratio, $g_e$, for 
of the electron's orbital motion even though $g_e=1$, just to 
keep track of its appearance. The energy for the interaction of the 
electron with the magnetic field now likewise receives a 
factor of $\gamma$.

Pauli now applies all this to the ``core model'' for atoms with 
a single valence electron.\footnote{Instead of the more modern 
expression ``valence electron'' Pauli speaks of ``light electron'' 
(German original: Lichtelektron). Sometimes the term ``radiating electron'' 
is also used (e.g., in \cite{Tomonaga:StoryOfSpin}).} 
According to the simplest version of this model, the total angular 
momentum, $\vec J$, and the total 
magnetic moment, $\vec M$, are the vector sums of the angular 
and magnetic momenta of the core (indicated here by a subscript 
$c$) and the valence electron (indicated here by a subscript $e$):
\begin{subequations}
\label{eq:TotalMomenta}
\begin{alignat}{2}
\label{eq:TotalMomenta-a}
&\vec J&&\,=\,\vec J_c+\vec J_e\,,\\
\label{eq:TotalMomenta-b}
 &\vec M&&\,=\,\vec M_c+\vec M_e\,. 
\end{alignat}
\end{subequations}
The relations between the core's and electron's magnetic momenta 
on one side, and their angular momenta on the other, are of the 
form 
\begin{subequations}
\label{eq:RelationsMomenta}
\begin{alignat}{2}
\label{eq:RelationsMomenta-a}
&\vec M_c&&\,=\,\frac{eg_c}{2m_0}\,\vec J_c\,,\\
\label{eq:RelationsMomenta-b}
&\vec M_e&&\,=\,\frac{eg_e}{2m_0}\,\vec J_e\,.
\end{alignat}
\end{subequations}

The point is now that $\vec M$ is not a multiple of $\vec J$ 
if $g_e\ne g_c$. Assuming a constant $\vec J$ for the time being, 
this means that $\vec M$ will precess around $\vec J$. Hence $\vec M$ 
is the sum  of a time independent part, $\vec M_\Vert$, parallel 
to $\vec J$ and a rotating part, $\vec M_\perp$, perpendicular 
to $\vec J$. The time average of $\vec M_\perp$ vanishes so that 
the effective magnetic moment is just given by $\vec M_\Vert$. 
Using (\ref{eq:TotalMomenta}) and (\ref{eq:RelationsMomenta}), 
and resolving scalar products into sums and differences of 
squares,\footnote{Like, e.g., 
$\vec J\cdot\vec J_e
=-\frac{1}{2}\bigl((\vec J-\vec J_e)^2-J^2-J_e^2\bigr)
=-\frac{1}{2}\bigl(J_c^2-J^2-J_e^2\bigr)$.}
we get
\begin{equation}
\label{eq:EffMagnMoment}
\begin{split}
\vec M_\Vert
&=\frac{\vec J\cdot\vec M}{J^2}\ \vec J\\
&=\frac{e}{2m_0}\
  \frac{g_e(\vec J\cdot\vec J_e)+g_c(\vec J\cdot\vec J_c)}{J^2}\ \vec J\\
&=\frac{e}{2m_0}\
  \frac{g_e(J^2+J_e^2-J_c^2)+g_c(J^2+J_c^2-J_e^2)}{2J^2}\ \vec J\\
&=\frac{e}{2m_0}\
   \left\{g_e+\bigl(g_c-g_e\bigr)\frac{J^2+J_c^2-J_e^2}{2J^2}\right\}\ \vec J\,.\\
\end{split}  
\end{equation}
Setting again $g_e=1$, the expression in curly brackets gives the 
gyromagnetic factor of the total system with respect to the effective 
magnetic moment. Its quantum analog is obtained by replacing 
$J^2\rightarrow J(J+1)$ and correspondingly for $J^2_c$ and $J^2_e$, which 
is then called the \emph{Land\'e factor} $g_L$. Hence 
\begin{equation}
\label{eq:LandeFactor}
g_L:=1+(g_c-1)\,\frac{J(J+1)+J_c(J_c+1)-J_e(J_e+1)}{2J(J+1)}\,.  
\end{equation}
All this is still right to a good approximation if $\vec J$ is not 
constant, but if its frequency of precession around the direction 
of the (homogeneous) external field is much smaller than the 
precession frequency of $\vec M$ around $\vec J$, which is the 
case for sufficiently small external field strength .

Basically through the work of Land\'e  it was known that $g_c=2$ 
fitted the observed multiplets of alkalies and also earth alkalies 
quite well. This value clearly had to be considered anomalous, since 
the magnetic moment and angular momentum of the core were due to the 
orbital motions of the electrons inside the core, which inevitably 
would lead to $g_c=1$, as explained in section\,\ref{sec:Preliminaries}. 
This was a great difficulty for 
the core model at the time, which was generally referred to as the 
``magneto-mechanical anomaly''. Pauli pointed out that one could 
either say that the physical value of the core's gyromagnetic 
factor is twice the normal value, or, alternatively, that it is 
obtained by adding 1 to the normal value. 

These two ways of looking at the anomaly  suggested two different 
ways to account for the relativistic correction, which should only affect 
that part of the magnetic moment that is due to the orbital motion 
of the inner electrons, that is, the `normal' part of 
$g_c$. Hence Pauli considered the following two possibilities 
for a relativistic correction of $g_c$, corresponding to the 
two views just outlined:
\begin{equation}
\label{eq:CorrectionsForG_c}
g_c=2\cdot 1\ \rightarrow\ g_c=2\cdot\gamma\qquad\text{or}\qquad
g_c=1+1\ \rightarrow\ g_c=1+\gamma\,.
\end{equation}

Then comes his final observation, that neither of these 
corrections are compatible with experimental results 
on high-$Z$ elements by Runge, Paschen and Back, which,
like the low-$Z$ experiments, resulted in compatibility 
with (\ref{eq:LandeFactor}) only if $g_c=2$. In a footnote 
Pauli thanked Land\'e and Back for reassuring him that 
the accuracy of these measurements where about one percent.
Pauli summarises his findings as follows
\begin{quote}
If one wishes to keep the hypothesis that the magneto-mechanical 
anomaly is also based in closed electron groups and, in particular, 
the $K$ shell, then it is not sufficient to assume a doubling of the 
ratio of the group's magnetic moment to its angular momentum relative 
to its classical value. In addition, one also needs to assume a 
compensation of the relativistic correction. 
(\cite{Pauli:CSP}, Vol.\,2, p.\,211)
\end{quote}
After some further discussion, in which he stresses once more 
the strangeness\footnote{For example: how can one understand 
the sudden doubling that the gyromagnetic factor of an outer 
electron  must suffer when joining the core?} that lies in 
$g_c=2$, he launches the following hypothesis, which forms 
the main result of his paper: 
\begin{quote}
The closed electron configurations shall not contribute to the magnetic
moment and angular momentum of the atom. In particular, for the alkalies,
the angular momenta of, and energy changes suffered by, the atom in an 
external magnetic field shall be viewed exclusively as an effect of the 
light-electron, which is also regarded as the location [``der Sitz''] of the 
magneto-mechanical anomaly. The doublet structure of the alkali spectra,
as well as the violation of the Larmor theorem, is, according to this 
viewpoint, a result of a classically indescribable two-valuedness of the 
quantum-theoretic properties of the light-electron.
(\cite{Pauli:CSP}, Vol.\,2, p.\,212)
\end{quote}
Note that this hypothesis replaces the atom's core as carrier of 
angular momentum by the valence electron. This means that 
(\ref{eq:TotalMomenta}), (\ref{eq:RelationsMomenta}), and 
(\ref{eq:LandeFactor}) are still valid, except that the subscript 
$c$ (for ``core'') is now replaced by the subscript $s$ (for ``spin'',
anticipating its later interpretation), so that we now have a 
coupling of the electron's orbital angular momentum (subscript 
$e$) to its intrinsic angular momentum (subscript $s$). 
In (\ref{eq:LandeFactor}), with $g_c$ replaced by $g_s$, one 
needs to set $g_s=2$ in order to fit the data. But now, as long 
as no attempt is made to relate the intrinsic angular momentum 
and magnetic moment of the electron to a common origin, there 
is no immediate urge left to regard this value as anomalous. Also, 
the problem in connection with the relativistic corrections 
(\ref{eq:CorrectionsForG_c}) now simply disappeared, since it was based 
on the assumption that $\vec J_c$ and $\vec M_c$ were due to orbital 
motions of inner (and hence fast) electrons, whereas in the new 
interpretation only $\vec J_e$ and $\vec M_e$ are due to orbital 
motion of the outer (and hence slow) valence electron. 

It is understandable that this hypothesis was nevertheless felt 
by some to lack precisely that kind of `explanation' that Pauli 
deliberately stayed away from: a common dynamical origin of 
the electron's inner angular momentum and magnetic moment.
From here the `story of spin' takes its course, leading to the 
hypothesis of the rotating electron, first conceived by Kronig 
and a little later, and apparently independently, by Goudsmit and 
Uhlenbeck, and finally to its implementation into Quantum Mechanics 
by Pauli~\cite{Pauli:1927} (``Pauli Equation'' for the non-relativistic 
case) and Dirac~\cite{Dirac:1928} (fully Lorentz invariant ``Dirac 
Equation''). Since then many myths surrounding spin built up, like that 
the concept of spin, and in particular the value $g=2$, was 
irreconcilable with classical (i.e. non-quantum) physics and that 
only the Dirac equation naturally predicted $g=2$. As for the latter
statement, it is well known that the principle of minimal coupling 
applied to the Pauli equation leads just as natural to $g=2$ as in case  
of the Dirac equation (cf. \cite{Galindo.Sanchez:1983} and 
\cite{Feynman:Electrodynamics}, p.\,37). Also, the very concept of spin 
has as natural a home in classical physics as in quantum physics if one 
starts from equally general and corresponding group-theoretic 
considerations.\footnote{\label{foot:ClassicalSpin}%
The spaces of states in quantum and classical mechanics are Hilbert 
spaces and symplectic manifolds respectively. An \emph{elementary system} 
is characterised in Quantum Mechanics by the requirement that the 
group of space-time symmetries act unitarily and irreducibly on its 
space of states. The corresponding requirement in Classical Mechanics 
is that the group action be symplectic and transitive~\cite{Bacry:1967}. 
The classification of homogeneous (with respect to the space-time 
symmetry group, be it the Galilei or Lorentz group) symplectic 
manifolds \cite{Arens:1971a}\cite{Guillemin.Sternberg:1990} leads 
then as natural to a classical concept of spin as the classification 
of unitary irreducible (ray-) representations leads to the 
quantum-mechanical spin concept. The mentioned classical structures 
are related to the quantum structures by various concepts of 
`quantisation' like `geometric quantisation'. Compare 
\cite{Woodhouse:GQ}, in particular Chap.\,6 on elementary systems.}

For the rest of this contribution I wish to concentrate on the particular 
side aspect already outlined in the introduction. Let me repeat the 
question: In what sense do the actual values of the electron parameters, 
mass, charge, intrinsic angular-momentum, and gyromagnetic factor, resist 
classical modelling in the framework 
of Special Relativity?

\section{Simple models of the electron}
\label{sec:ElectronModels}
In this section we will give a self-contained summary of the basic 
features of simple electron models. The first model corresponds 
to that developed by Abraham~\cite{Abraham:1902}, which was 
mentioned by Goudsmit and Uhlenbeck as already 
explained.\footnote{\label{foot:AbrahamLorentz}% 
Since we are mainly concerned with the spin aspects, we will ignore 
the differences between Abraham's and, say, Lorentz' model (rigid 
versus deformable), which become important as soon as translational 
motions are considered. We mention Abraham not for any preference 
for his `rigid' model, but for the reason that he considered rotational 
motion explicitly. Its interaction with the translational motion was 
further worked out in detail by Schwarzschild in~\cite{Schwarzschild:1903c}, 
but this is not important here.} We will see that this model can 
only account for $g$ factors in the interval between $3/2$ and $11/6$ 
if superluminal speeds along the equator are to be avoided. We also 
critically discuss the assumption made by Goudsmit and Uhlenbeck that 
this (i.e. Abraham's) model predicts $g=2$. Since this model neglects 
the stresses that are necessary to prevent the charge distribution 
from exploding, we also discuss a second model in which such stresses 
(corresponding to a negative pressure in the electron's interior) are 
taken into account, at least in some slow-rotation approximation. 
This model, too, has been discussed in the literature 
before~\cite{Cohen.Mustafa:1986a}. Here it is interesting to see 
that due to those stresses significantly higher values of $g$ are 
possible, though not for small charges as we will also 
show.\footnote{This is another example of a special-relativistic 
effect which has nothing to do with large velocities.} 
Finally we discuss the restriction imposed by the condition of 
energy dominance, which basically says that the speed of sound of 
the stress-supporting material should not exceed the speed 
of light. This sets an upper bound on $g$ given by $9/4$. 
Note that all these statements are made only in the realm where 
the slow-rotation approximation is valid. I do not know of any fully 
special-relativistic treatment on which generalisations of these 
statements could be based. In that sense, the general answer to 
our main question posed above is still lacking.

\subsection{A purely electromagnetic electron}
\label{sec:ElectronModelsAbraham}
Consider a homogeneous charge distribution, $\rho$, of total charge $Q$ on 
a sphere of radius $R$ centred at the origin (again we write 
$r:=\Vert\vec x\Vert$ and $\vec n:=\vec x/r$):
\begin{equation}
\label{eq:ChargeDensity}
\rho(\vec x)
=\frac{Q}{4\pi R^2}\,\delta(r-R)\,.
\end{equation}
For the moment we shall neglect the rest mass of the matter that sits at 
$r=R$ and also the stresses it must support in order to keep the charge 
distribution in place. The charge is the source of the scalar potential 
\begin{equation}
\label{eq:ScalarPotential}
\phi(\vec x)\
=\frac{1}{4\pi\varepsilon_0}
 \int\frac{\rho(\vec x')}{\Vert\vec x-\vec x'\Vert}\,d^3x'
=\frac{Q}{4\pi\varepsilon_0\,R}
\begin{cases}
1   & \text{for $r<R$}\,,\\
R/r & \text{for $r>R$}\,,
\end{cases}
\end{equation}
with corresponding electric field

\begin{equation}
\label{eq:ElectricField}
\vec E(\vec x)=
\frac{Q}{4\pi\varepsilon_0\,R^2}
\begin{cases}
\vec 0   & \text{for $r<R$}\,,\\
\vec n   & \text{for $r>R$}\,.
\end{cases}
\end{equation}
Let now the charge distribution rotate rigidly with constant 
angular velocity $\vec\omega$. This gives rise to a current 
density 
\begin{equation}
\label{eq:CurrentDensity}
\vec j(\vec x)
=(\vec\omega\times\vec x)\,\rho(\vec x)
=\frac{Q}{4\pi R^2}\,(\vec\omega\times\vec x)\,\delta(r-R)\,,
\end{equation}
which, in turn, is the source of a vector potential according to
\begin{equation}
\label{eq:BiotSavartsLaw1}
\vec A(\vec x)
=\frac{\mu_0}{4\pi}\int\frac{\vec j(\vec x')}{\Vert\vec x-\vec x'\Vert}\,d^3x'
=\frac{\mu_0 Q}{12\pi R}\,\vec\omega\times
\begin{cases}
\vec x & \text{for $r<R$}\,,\\
\vec x\,(R/r)^3 & \text{for $r>R$}\,.
\end{cases}
\end{equation}
Hence, in the rotating case, there is an additional magnetic field 
in addition to the electric field~(\ref{eq:ElectricField}):
\begin{equation}
\label{eq:BiotSavartsLaw2}
\vec B(\vec x)=\frac{\mu_0}{4\pi}
\begin{cases}
2\vec M/R^3 & \text{for $r<R$}\,,\\
\bigl(3\vec n(\vec n\cdot\vec M)-\vec M\bigr)/r^3 & \text{for $r>R$}\,,
\end{cases}
\end{equation}
where
\begin{equation}
\label{eq:MagnDipoleMoment}
\vec M:=\tfrac{1}{3}QR^2\,\vec\omega\,.
\end{equation}
For $r<R$ this is a constant field in $\vec\omega$ direction. 
For $r>R$ it is a pure dipole field (i.e. all higher 
multipole components vanish) with moment 
(\ref{eq:MagnDipoleMoment}).

\subsubsection{Energy}
The general expression for the energy of the electromagnetic field 
is\footnote{From now on we shall denote the modulus of a vector 
simply by its core symbol, i.e., $\Vert\vec E\Vert=E$ etc.} 
\begin{equation}
\label{eq:TotalFieldEnergy1}
\mathcal{E}=\int_{\mathbb{R}^3}\tfrac{1}{2}
\Bigl(\varepsilon_0 E^2(\vec x)
+\tfrac{1}{\mu_0} B^2(\vec x)\Bigr)\,d^3x\,.
\end{equation}
For the case at hand, the electric and magnetic contributions 
to the energy are respectively given by
\begin{subequations}
\label{eq:Electric-MagneticFieldEnergy}
\begin{alignat}{3}
\label{eq:Electric-MagneticFieldEnergy-a}
& \mathcal{E}_e&&=\quad\frac{Q^2}{8\pi\varepsilon_0\,R}
&& \begin{cases}
0 & \text{from $r<R$}\\
1 & \text{from $r>R$}
\end{cases}\\
&\mathcal{E}_m&&\,=\,\frac{\mu_0}{4\pi}\,M^2/R^3\,
&&\begin{cases}
2/3 & \text{from $r<R$}\\
1/3 & \text{from $r>R$}\,.
\end{cases}
\end{alignat}
\end{subequations}
The total magnetic contribution can be written as
\begin{equation}
\label{eq:Total MagneticFieldEnergy}
\mathcal{E}_m=\frac{\mu_0}{4\pi}\,M^2/R^3
=\tfrac{1}{2}\,I\,\omega^2\,,
\end{equation}
where
\begin{equation}
\label{eq:EmMomentInertia}
I:=\frac{\mu_0}{18\pi}\,Q^2R
\end{equation}
may be called the \emph{electromagnetic moment of inertia}~\cite{Abraham:1902}.
It has no mechanical interpretation in terms of a rigid rotation 
of the electrostatic energy distribution (see below)!

The total electromagnetic energy can now be written as 
\begin{equation}
\label{eq:TotalFieldEnergy2}
\mathcal{E}=\mathcal{E}_e+\mathcal{E}_m
=\frac{Q^2}{8\pi\varepsilon_0\,R}
\Bigl\{1+\tfrac{2}{9}\beta^2\Bigr\}\,,
\end{equation}
where we used $\varepsilon_0\mu_0=1/c^2$ and set $\beta:=R\omega/c$.
The ratio of magnetic (`kinetic') to total energy is then given by 
\begin{equation}
\label{eq:FieldEnergyRatio}
\frac{\mathcal{E}_m}{\mathcal{E}}=\frac{\beta^2}{9/2+\beta^2}\,,
\end{equation}
which is a strictly monotonic function of $\beta$ bounded above by 1 
(as it should be). However, if we require $\beta<1$, the upper bound 
is $2/11$.

\subsubsection{Angular momentum}
\label{sec:AngularMomentumFirstModel}
The momentum density of the electromagnetic field vanishes for 
$r<R$ and is given by 
\begin{equation}
\label{eq:MomentumDensity}
\vec p(\vec x)=\frac{\mu_0}{16\pi^2}
\,Q\,(\vec M\times\vec n)/r^5
\end{equation}
for $r>R$ ($1/c^2$ times `Poynting vector'). 
The angular-momentum density also vanishes for $r<R$. For 
$r>R$ it is given by 
\begin{equation}
\label{eq:AngularMomentumDensity}
\vec \ell(\vec x)=\vec x\times\vec p(\vec x)=
\frac{\mu_0}{16\pi^2}\,Q\,\frac{\vec M-\vec n(\vec n\cdot\vec M)}{r^4}\,.
\end{equation}
Hence the total linear momentum vanishes, whereas the total 
angular momentum is given by 
\begin{equation}
\label{eq:TotalAngularMomentumMod1}
\vec J:=\int_{r>R}\vec\ell(\vec x)\,d^3x=I\vec\omega
\end{equation}
with the \emph{same} $I$ (moment of inertia) as in 
(\ref{eq:EmMomentInertia}).

\subsubsection{The gyromagnetic factor}
The gyromagnetic ratio now follows from expressions 
(\ref{eq:MagnDipoleMoment}) for $\vec M$ and 
(\ref{eq:TotalAngularMomentumMod1}) for $\vec J$:
\begin{equation}
\label{eq:GyromagnRatioEM}
\frac{M}{J}
=\frac{6\pi\,R}{\mu_0\,Q}=:g\,\frac{Q}{2m}\,,
\end{equation}
where $m$ denotes the total mass, which is here given by 
\begin{equation}
\label{eq:Total Mass}
m:=\mathcal{E}/c^2
=\frac{\mu_0}{8\pi}\frac{Q^2}{R}\Bigl\{1+\tfrac{2}{9}\beta^2\Bigr\}\,.
\end{equation}
Hence $g$ can be solved for:
\begin{equation}
\label{eq:EM-gFactor}
g=\frac{3}{2}\,\Bigl\{1+\tfrac{2}{9}\beta^2\Bigr\}\,,
\end{equation}
so that 
\begin{equation}
\label{eq:EM-gFactorRange}
\tfrac{3}{2}<g<\tfrac{11}{6}\qquad\text{if}\qquad
0<\beta<1\,.
\end{equation}
Even with that simple model we do get quite close to $g=2$.

\subsubsection{Predicting $g=2$?}
It is sometimes stated that Abraham's model somehow `predicts' $g=2$ 
(e.g., \cite{Pais:1989} p.\,39 or \cite{Pfister.King:2003} p.\,206),
though this is not at all obvious from \cite{Abraham:1902}. 
My interpretation for how such a `prediction' could come about can 
be given in terms of the present special-relativistic model.\footnote{%
Here we ignore Abraham's rigidity condition which would complicate the 
formulae without changing the argument proper. Also recall  
footnote\,\ref{foot:AbrahamLorentz}.} It rests on an 
(inconsistent) combination of the following two observations. First, 
if we Lorentz transform the purely electric field (\ref{eq:ElectricField}) 
into constant translational motion with velocity $w$, we obtain a 
new electric and also a non-vanishing magnetic field. The integrated 
Poynting  vector then gives the total electromagnetic momentum of the 
charged shell at speed $w$:
\begin{equation}
\label{eq:EmMomentum4/3}
p=\frac{4}{3}\frac{m_ew}{\sqrt{1-w^2/c^2}}\,,
\end{equation}
where 
\begin{equation}
\label{eq:ElMass}
m_e:=\mathcal{E}_e/c^2=\frac{\mu_0}{8\pi}\frac{Q^2}{R}\,.
\end{equation}
The infamous factor 4/3 results from the contribution of the (unbalanced) 
electromagnetic stresses.\footnote{\label{foot:FourThirds} Generally 
speaking, the factor 4/3 marks the discrepancy between two definitions 
of `electromagnetic mass', one through the electromagnetic momentum, 
the other, called $m_e$ above, through the electrostatic energy. This 
discrepancy is nothing to get terribly excited about and simply a 
consequence of the non-conservation of the electromagnetic 
energy-momentum tensor, i.e., $\nabla_\mu T_{em}^{\mu\nu}\ne 0$, 
a result of which is that the (unbalanced) electromagnetic stresses 
contribute to 
the electromagnetic momentum another third of the expression
$p=m_ew/\sqrt{1-w^2/c^2}$ that one naively obtains from 
just formally transforming total energy and momentum as 
time and space components respectively of a four vector. 
Much discussion in the literature was provoked by getting 
confused whether this state of affairs had anything to do 
with Lorentz non-covariance. See, e.g., 
\cite{Campos.Jimenez:1986} for a good account and 
references.}  In this way one is led to assign to the electron 
a dynamically measurable rest-mass of $m=\frac{4}{3}m_e$ 
\emph{if one neglects the rotational energy}. Second, we 
may ask how fast the electron is to spin for (\ref{eq:Total Mass}) 
to just give $m=\frac{4}{3}m_e$ (rest energy of the spinning electron). 
The immediate answer is, that this is just the case if 
and only if $1+\frac{2}{9}\beta^2=\frac{4}{3}$, which in 
view of (\ref{eq:EM-gFactor}) is equivalent to $g=2$. 

It is now obvious how this argument rests on the conflation 
of two different notions of mass. The factor $4/3$ will  
consistently be dealt with by taking into account the 
stresses that balance electrostatic repulsion, not by 
trying to account for it in letting the electron spin fast 
enough.

\subsection{A side remark on the kinematics of Faraday lines}
\label{sec:KinematicsFaradayLines}
In the Introduction we stressed that the emancipation of the notion 
of angular momentum from the usual kinematical notion of rotation 
in space had already begun in classical field theory. More precisely
this applies to Maxwell's theory, in which the notion of a field 
differs from that of, say, hydrodynamics in that it is \emph{not} 
thought of as being attached to a material carrier. This has 
consequences if we wish to apply kinematical states of motion 
to the field itself. 

At first sight, Faraday's picture of lines of force in space suggests 
to view them as material entities, capable of assuming different 
kinematical states of motion. If so, the time-dependence of the 
electromagnetic field might then be interpreted as, and possibly 
explained by, the motions of such lines (given by some yet unknown 
equations of motion, of which the Maxwell equations might turn out 
to be some coarse grained version). That this is not possible has 
been stressed by Einstein in his 1920 Leiden address ``Ether and 
the Theory of Relativity'', where he writes 
\begin{quote}
If one wishes to represent these lines of force as something material
in the usual sense, one is tempted to interpret dynamical processes 
[of the em. field] as motions of these lines of force, so that each 
such line can be followed in time. It is, however, well known that 
such an interpretation leads to contradictions.\\ 
In general we have to say that it is possible to envisage extended 
physical objects to which the notion of motion [in space] does not  
apply. (\cite{Einstein:CP}, Vol.\,7, Doc.\,38, p.\,315)
\end{quote}   

The reason why we mention this is that the notion of an 
``electromagnetic moment of inertia'', 
introduced in (\ref{eq:EmMomentInertia}), nicely illustrates 
this point. Assume that the electrostatic energy density 
$\rho_e$ of the Coulomb field of charge $Q$ corresponded 
to a mass density according to a local version of 
$E=mc^2$, i.e., 
\begin{equation}
\label{eq:ElectrostaticMassDensity}
\rho_m(\vec x)
:=\rho_e(\vec x)/c^2=\left(\frac{\mu_0}{32\pi^2}\right)\frac{Q^2}{r^4}\,.
\end{equation}
If the electrostatic energy is now thought of as being attached to the 
somehow individuated lines of force, a moment of inertia for 
the shell $R<r<R'$ would result, given by 
\begin{equation}
\label{eq:ElectrostaticMoI}
I(R')
=\int_{R<r<R'}\rho_m(\vec x)\,(r\sin\theta)^2\,d^3x
=\left(\frac{2\mu_0}{27\pi}\right)\,Q^2\,(R'-R)\,.
\end{equation}
But this diverges as $R'\rightarrow\infty$, in contrast to 
(\ref{eq:EmMomentInertia}), showing that we may not think of 
the energy distribution of the electromagnetic field as 
rigidly rotating in the ordinary sense. 

\subsection{An electron model with Poincar\'e stresses}
\label{sec:ElectronModelsCohen}
In this section we will modify the previous model for the electron 
in the following three aspects
\begin{itemize}
\item[1.]
The infinitesimally thin spherical shell is given a small rest-mass
of constant surface density $m_0/4\pi R^2$.
\item[2.]
Stresses in the shell are taken into account which keep the electron
from exploding. They are called ``Poincar\'e stresses'' since 
Poincar\'e was the first in 1906 to discuss the dynamical need 
of balancing stresses~\cite{Poincare:1906}\cite{Miller:1973}.
\item[3.]
The rotational velocity is small, so that $(R\omega/c)^n$ terms are 
neglected for $n\geq 2$.
\end{itemize}

\subsubsection{Poincar\'e stress}
\label{sec:PoincareStress}
The second modification needs further explanation.
If we view the surface $r=R$ as a kind of elastic membrane, there 
will be tangential stresses in the surface of that membrane that 
keep the charged membrane from exploding. In the present approximation,
which keeps only linear terms in $\omega$, these stresses need only 
balance the electrostatic repulsion, which is constant over the 
surface $r=R$. In quadratic order the stresses would, in addition, 
need to balance the latitude dependent centrifugal forces, which we 
neglect here. 

To calculate the surface stress that is needed to balance electrostatic 
repulsion we recall the expression (\ref{eq:Electric-MagneticFieldEnergy-a})
for the electrostatic energy as function of radius $R$:
\begin{equation}
\label{eq:ElectricFieldEnergy}
\mathcal{E}_e=\quad\frac{Q^2}{8\pi\varepsilon_0\,R}\,.
\end{equation}
Varying $R$ gives us the differential of work that we need to 
supply in order to change the volume through a variation of $R$.
Equating this to $-pdV=-p\,4\pi R^2\,dR$ gives the pressure 
inside the electron:
\begin{equation}
\label{eq:ElectronPressure}
p=\left(\frac{1}{4\pi\varepsilon_0}\right)\frac{Q^2}{8\pi\,R^4}\,.
\end{equation}
Now, imagine the sphere $r=R$ being cut into two hemispheres along 
a great circle. The pressure tries to separate these hemispheres 
by acting on each with a total force of strength $p\pi R^2$ in 
diametrically opposite directions.\footnote{This follows 
immediately from the general fact that the total force along 
a given direction that a constant pressure exerts on a surface 
is given by the pressure times the area of the planar projection 
of that surface perpendicular to the given direction. 
Alternatively it may be verified directly through integrating 
the element of force in polar direction (i.e. perpendicular 
to the surface spanned by the great circle), 
$dF=(p\cos\theta)(R^2\sin\theta d\theta d\varphi)$, 
over a hemisphere.} This force is distributed uniformly 
along the cut (the great circle), whose length is $2\pi R$. Hence the 
force per length is just $pR/2$. The surface stress, $\sigma$, 
(force per length) that is needed to prevent the electron from 
exploding is just the negative of that. Using 
(\ref{eq:ElectronPressure}), we therefore get
\begin{equation}
\label{eq:SurfaceStress}
\sigma=\,-\,\left(\frac{1}{4\pi\varepsilon_0}\right)
\frac{Q^2}{16\pi R^3}\,.
\end{equation}

\subsubsection{Energy-momentum tensor}
\label{sec:EMT}
The energy-momentum tensor now receives a contribution that accounts 
for the presence of the surface stress (\ref{eq:SurfaceStress}) that 
acts tangential to the surface $r=R$ \emph{in the local rest frame 
corresponding to each surface element of the rotating sphere}. The four-velocity 
of each surface element is given by\footnote{I use spacetime 
coordinates $(t,r,\theta,\varphi)$ where the latter three are standard 
spherical polar coordinates. I also employ the notation 
$\partial_\mu:=\partial/\partial x^\mu$ for the chart-induced 
vector fields, so that, e.g., $\partial_\varphi:=\partial/\partial\varphi$.}
\begin{equation}
\label{eq:FourVel}
u=\partial_t+\omega\,\partial_{\varphi}\,,
\end{equation}
which is  normalised ($g(u,u)=c^2$) up to terms 
$\omega^2$ (which we neglect). Recall that the space-time metric of 
Minkowski space in spatial polar coordinates is (we use the 
``mostly plus'' convention for the signature) 
\begin{equation}
\label{eq:MinkMetric}
g=
%g_{\mu\nu}dx^\mu\otimes dx^\nu
-\,c^2\,dt\otimes dt+dr\otimes dr+r^2\,d\theta\otimes d\theta+
r^2\sin^2\theta\,d\varphi\otimes d\varphi\,.
\end{equation}

The energy-momentum tensor has now three contributions,
corresponding to the matter of the shell (subscript $m$), the 
Poincar\'e stresses within the shell (subscript $\sigma$), 
and the electromagnetic field (subscript $em$):
\begin{subequations}
\label{eq:FullEM-Tensor}
\begin{equation}
\label{eq:FullEM-Tensor-a}
\mathbf{T} =
\mathbf{T}_{m}+
\mathbf{T}_{\sigma}+
\mathbf{T}_{em}\,.
\end{equation}
The first two comprise the shell's contribution and are given by
\begin{alignat}{2}
\label{eq:FullEM-Tensor-b}
& \mathbf{T}_m
&&\,=\,\frac{m_0}{4\pi R^2}\,\delta(r-R)\,u\otimes u\,,\\
\label{eq:FullEM-Tensor-c}
& \mathbf{T}_\sigma
&&\,=\,\,-\, \left(\frac{1}{4\pi\varepsilon_0}\right)
\frac{Q^2}{16\pi R^3}\,\delta(r-R)\,\mathbf{P}\,.
\end{alignat}
\end{subequations}
Here $\mathbf{P}$ is the orthogonal projector onto the 2-dimensional 
subspace orthogonal to $u$ and $\partial_r$, which is the subspace 
tangential to the sphere in each of its local rest frames. It can be 
written explicitly in terms of local orthonormal 2-legs, $n_1$ and 
$n_2$, spanning these local 2-planes. For example, we may take 
$n_1:=\frac{1}{r}\partial_\theta$ and write (since $n_2$ must be 
orthogonal to $\partial_r$ and $\partial_\theta$) 
$n_2=a\partial_t+b\partial_\varphi$, where the coefficients $a,b$ 
follow from $g(u,n_2)=0$ and normality. This gives  
\begin{subequations}
\label{eq:Projector}
\begin{equation}
\label{eq:Projector-a}
\mathbf{P}=n_1\otimes n_1+n_2\otimes n_2\,,
\end{equation}
where 
\begin{alignat}{2}
\label{eq:Projector-b}
&n_1&&\,:=\,\tfrac{1}{r^2}\,\partial_\theta\,,\\
\label{eq:Projector-c}
&n_2&&\,:=\,c^{-2}\omega\,r\sin\theta\,\partial_t
+(r\sin\theta)^{-1}\partial_\varphi\,.
\end{alignat}
\end{subequations}
Note that $g(n_1,n_1)=g(n_2,n_2)=1$ and $g(n_1,n_2)=0$. 
Equation (\ref{eq:Projector-a}) may therefore be be written 
in the form (again neglecting $\omega^2$ terms)
\begin{equation}
\label{eq:Projector-Alt}
\mathbf{P}
=r^{-2}\,\partial_\theta\otimes\partial_\theta
+(r\sin\theta)^{-2}\,\partial_\varphi\otimes\partial_\varphi
+c^{-2}\omega\,\bigl(
\partial_t\otimes\partial_\varphi+\partial_\varphi\otimes\partial_t
\bigr)\,.
\end{equation}
For us the crucial term will be the last one, which is off-diagonal,
since it will contribute to the total angular momentum. More precisely, 
we will need to invoke the integral of 
$\partial_t\cdot\mathbf{P}\cdot\partial_\varphi)$ (the dot ($\,\cdot\,$)  refers 
to the inner product with respect to the Minkowski metric) 
over the sphere $r=R$:
\begin{equation}
\label{eq:IntP}
\int(\partial_t\cdot\mathbf{P}\cdot\partial_\varphi)\,
R^2\sin\theta d\theta d\varphi
=\int c^{-2}\,\omega g_{tt}g_{\varphi\varphi}\,
R^2\sin\theta d\theta d\varphi
=-\,\tfrac{8\pi}{3}\omega R^4\,.
\end{equation}
where we used $g_{tt}:=g(\partial_t,\partial_t)=-c^2$ and 
$g_{\varphi\varphi}:=g(\partial_\varphi,\partial_\varphi)
=R^2\sin^2\theta$ from (\ref{eq:MinkMetric}).

\subsubsection{A note on linear momentum and von\,Laue's theorem}
\label{sec:LinearMomentum}
The addition of the stress part has the effect that the total 
energy-momentum tensor is now conserved (here in the slow-rotation
approximation):
\begin{equation}
\label{eq:EM-Conservation}
\nabla_\mu T^{\mu\nu}=0\,,
\end{equation}
as one may explicitly check. Note that since we use curvilinear
coordinates here we need to invoke the covariant 
derivative.\footnote{We have$
\nabla_\mu T^{\mu\nu}
=\partial_\mu T^{\mu\nu}
+\Gamma_{\mu\lambda}^{\mu}T^{\lambda\nu}
+\Gamma_{\mu\lambda}^{\nu}T^{\mu\lambda}$, 
where $\Gamma_{\mu\lambda}^{\nu}:=\frac{1}{2}g^{\nu\sigma}\bigl(
-\partial_\sigma g_{\mu\lambda}
+\partial_\lambda g_{\sigma\mu}
+\partial_\mu g_{\lambda\sigma}\bigr)$, with $g_{\mu\nu}$ taken from
(\ref{eq:MinkMetric}). The $\Gamma$'s are most easily computed 
directly from the geodesic equation.}
Indeed, writing the shell's energy momentum tensor as 
$\mathbf{T}_s:=\mathbf{T}_m+\mathbf{T}_\sigma$, it is not 
difficult to show that $\nabla_\mu T_s^{\mu\nu}$ is zero for 
$\nu\ne r$, and for $\nu=r$ is given by $p\,\delta(r-R)$ with 
$p$ as in  (\ref{eq:ElectronPressure}). But this clearly equals 
$-\nabla_\mu T_{em}^{\mu\nu}$ since, according to Maxwell's 
equations, this quantity equals minus the electromagnetic 
force density on the charge distribution, which is obviously 
$-p\,\delta(r-R)$. In fact, this is precisely the interpretation 
that we used to determine $p$ in the first place.

The conservation equation (\ref{eq:EM-Conservation}) generally 
ensures that total energy and total momentum form, respectively, 
the time- and space component of a four vector. Let us now
show explicitly that $T_\sigma$ removes the factor $4/3$ in 
the calculation of the linear momentum when the system is boosted 
in ,say, the $z$ direction. To do this we need to calculate the 
integral of $\partial_z\cdot\mathbf{T}_\sigma\cdot\partial_z$ 
over all of space and show that it precisely cancels the 
corresponding integral of the electromagnetic part, i.e. the 
integral over $\partial_z\cdot\mathbf{T}_{em}\cdot\partial_z$. 
Noting that $g(\partial_\theta,\partial_z)=r\sin\theta$, we 
have 
\begin{equation}
\label{eq:VerLaueTheorem-1}
\int dV\,\bigl(\partial_z\cdot\mathbf{T}_\sigma\cdot\partial_z\bigr)
=\int drd\theta\, d\varphi\bigl(\sigma\,\delta(r-R)\,r^2\sin^3\theta\bigr)
=\tfrac{8\pi}{3}\sigma R^2=-\tfrac{1}{3}\mathcal{E}_e\,,
\end{equation}
whereas the tracelessness of $\mathbf{T}_{em}$ together with isotropy 
immediately imply  
\begin{equation}
\label{eq:VerLaueTheorem-2}
\int dV\,\bigl(\partial_z\cdot\mathbf{T}_{em}\cdot\partial_z\bigr)
=\tfrac{1}{3}\int dV\,
c^{-2}\bigl(\partial_t\cdot\mathbf{T}_{em}\cdot\partial_t\bigr)
=\tfrac{1}{3}\mathcal{E}_e\,.
\end{equation}
That the sum of (\ref{eq:VerLaueTheorem-1}) and (\ref{eq:VerLaueTheorem-2}) 
vanishes is a consequence of Laue's theorem, which basically states 
that the integral over all of space of the space-space components of a
time-independent conserved energy-momentum tensor vanish. 
Here this was achieved by including stresses, which subtracted one 
third of the electromagnetic linear momentum.\footnote{The requirement on 
the stress part $\mathbf{T}_\sigma$ to be such that the total energy 
and momentum derived from $\mathbf{T}_{em}+\mathbf{T}_\sigma$ should 
transform as a four vector clearly still leaves much freedom in the 
choice of $\mathbf{T}_\sigma$. The choice made here is such that the 
total rest energy equals the electrostatic self energy. But other 
values for the rest energy (like, e.g., 4/3 of the electrostatic 
contribution) would also have been possible. In particular, the 
`covariantisation through stresses' 
does not as such prefer any of the `electromagnetic masses' 
mentioned above (footnote\,\ref{foot:FourThirds}), as has also 
been demonstrated in an elegant and manifestly covariant fashion in 
\cite{Schwinger:1983}.} Similarly, the stresses will also subtract 
from the electromagnetic angular momentum, this time even the larger 
portion of three quarters of it. Moreover, since the magnetic moment 
is the same as before, the stresses will have the tendency to increase the 
gyromagnetic ratio. This we will see next in more detail   

\subsubsection{Angular momentum}
\label{sec:AngularMomentumSecondModel}
The total angular momentum represented by (\ref{eq:FullEM-Tensor})
is calculated by the general formula
\begin{subequations}
\label{eq:AngMom2Mod}
\begin{equation}
\label{eq:AngMom2Mod-a}
J=\,-\,\frac{1}{c^2}\,
\int\partial_t\cdot\mathbf{T}\cdot\partial_\varphi\,d^3x
=J_m+J_\sigma+J_{em}\,.
\end{equation}
The matter part, $J_m$, corresponding to (\ref{eq:FullEM-Tensor-b}),
yields the standard expression for a mass-shell of uniform density:
\begin{equation}
\label{eq:AngMom2Mod-b}
J_m=\tfrac{2}{3}m_0\omega R^2\,.
\end{equation} 
The electromagnetic part is the same as that already calculated,
since the electromagnetic field is the same. Therefore we just read 
off (\ref{eq:TotalAngularMomentumMod1}) and (\ref{eq:EmMomentInertia})
that 
\begin{equation}
\label{eq:AngMom2Mod-c}
J_{em}=\tfrac{2}{3}\cdot\tfrac{2}{3}\,m_e\omega R^2\,.
\end{equation} 
Finally, using (\ref{eq:IntP}), the contribution of the stresses 
can also be written down: 
\begin{equation}
\label{eq:AngMom2Mod-d}
J_\sigma
=\,-\,\tfrac{1}{2}\cdot\tfrac{2}{3}\,m_e\omega R^2
=\,-\,\tfrac{3}{4}\,J_{em}\,.
\end{equation} 
Adding the last two contributions shows that the inclusion of
stresses amounts to reducing the electromagnetic contribution 
from the value given by (\ref{eq:AngMom2Mod-b}) to a quarter 
of that value:
\begin{equation}
\label{eq:AngMom2Mod-e}
J_{em}+J_\sigma=J_{em}-\tfrac{3}{4}J_{em}=\tfrac{1}{4}J_{em}
\end{equation} 
In total we have  
\begin{equation}
\label{eq:TotAngMom2Mod-1}
J=\Bigl(m_0+\tfrac{1}{6}m_e\Bigr)\tfrac{2}{3}\omega R^3\,.
\end{equation}
\end{subequations} 

To linear order in $\omega$ the kinetic energy does not contribute to 
the overall mass, $m$, which is therefore simply given by 
the sum of the bare and the electrostatic mass
\begin{equation}
\label{eq:TotMass2Mod}
m=m_0+m_{e}\,.
\end{equation} 
Using this to eliminate $m_e$ in (\ref{eq:TotAngMom2Mod-1}) gives
\begin{equation}
\label{eq:TotAngMom2Mod-2}
J=\left(\frac{1+5m_0/m}{6}\right)\left(\,\frac{2}{3}m\omega R^2\right)\,.
\end{equation}

\subsubsection{The gyromagnetic factor}
\label{sec:GyrFac2Mod}
Since the electromagnetic field is exactly as in the previous model, 
the magnetic moment in the present case is that given by 
(\ref{eq:MagnDipoleMoment}). The gyromagnetic factor is defined 
through  
\begin{equation}
\label{eq:DefGyrFac2Mod}
\frac{M}{J}=g\frac{Q}{2m}\,,
\end{equation}
which leads to 
\begin{equation}
\label{eq:g-Fac2Mod}
g=\frac{6}{1+5m_0/m}\,.
\end{equation}
This allows for a range of $g$ given by
\begin{equation}
\label{eq:Rangeg-Fac2Mod}
1\leq g\leq 6\,,
\end{equation}
where $g=1$ corresponds to $m=m_0$, i.e., no electromagnetic 
contribution and $g=6$ corresponds to $m_0=0$, i.e., all mass is 
of electromagnetic origin. The interval (\ref{eq:Rangeg-Fac2Mod}) 
looks striking, given the modern experimental values for the 
electron and the proton:
\begin{equation}
\label{eq:g-FactorsElectronProton}
g_{\rm electron}=2.002 319 304 3622
\quad\text{and}\quad 
g_{\rm proton}=5.585 694 713\,.
\end{equation}
However, we have not yet discussed the restrictions
imposed by our slow-rotation assumption. This we shall do next.

\subsubsection{Restrictions by slow rotation}
\label{sec:RestSlowRot}
Our model depends on the four independent parameters, 
$P=(m_0,Q,R,\omega)$. On the other hand, there are four 
independent physical observables, $O=(m,Q,g,J)$ ($M$ is dependent 
through (\ref{eq:DefGyrFac2Mod})). 
Our model provides us with a functional dependence
expressing the observables as functions of the parameters: $O=O(P)$. 
Since $Q$ is already an observable, it remains to display $m,g,J$
in terms of the parameters:
\begin{subequations}
\label{eq:FunctObsPara}
\begin{alignat}{2}
\label{eq:FunctObsPara-a}
&m(m_0,Q,R)&&\,=\,m_0+\frac{\mu_0}{8\pi}\frac{Q^2}{R}=:m_0+m_e(Q,R)\,,\\
\label{eq:FunctObsPara-b}
&g(m_0,Q,R)&&\,=\,\frac{6}{1+5m_0/m(m_0,Q,R)}\,,\\
\label{eq:FunctObsPara-c}
&J(m_0,Q,R,\omega)&&\,=\,\Bigl(m_0+\tfrac{1}{6}m_e(Q,R)\Bigr)\,
\tfrac{2}{3}\omega R^2\,.
\end{alignat}
\end{subequations}
These relations can be inverted so as to allow the calculation of 
the values of the parameters from the values of the observables. 
If we choose to display $\beta:=R\omega/c$ rather than 
$\omega$, this gives
\begin{subequations}
\label{eq:FunctParaObs}
\begin{alignat}{2}
\label{eq:FunctParaObs-a}
&m_0(m,g)&&\,=\,m\,\frac{6-g}{5g}\,,\\
\label{eq:FunctParaObs-b}
&m_e(m,g)&&\,=\,m-m_0\,=\,m\,\frac{6(g-1)}{5g}\,,\\
\label{eq:FunctParaObs-c}
&R(m,Q,g)&&\,=\,\frac{\mu_0}{8\pi}\frac{Q^2}{m_e}\,=\,
                \frac{\mu_0}{8\pi}\frac{Q^2}{m}\frac{5g}{6(g-1)}\,\\
\label{eq:FunctParaObs-d}
&\beta(J,Q,g)&&\,=\,2J\,\left[\frac{Q^2}{4\pi\varepsilon_0 c}\right]^{-1}
                           \frac{9(g-1)}{5}\,,
\end{alignat}
\end{subequations}
where the last equation (\ref{eq:FunctParaObs-d}) follows from 
(\ref{eq:FunctObsPara-c}) using 
(\ref{eq:FunctParaObs-a}-\ref{eq:FunctParaObs-c}).
It is of particular interest to us since it allows to easily 
express the slow-rotation assumption $\beta\ll 1$. For this it will be 
convenient to measure $Q$ in units of the elementary charge $e$ 
and $J$ in units of $\hbar/2$. Hence we write 
\begin{equation}
\label{eq:UnitMeasures}
Q=n_Q\,e\qquad\text{and}\qquad 2J=n_J\,\hbar\,.
\end{equation}
Then, using that the fine-structure constant in SI units reads
$\alpha=e^2/(4\pi\varepsilon_0c\hbar)\approx 1/137$, we get 
\begin{equation}
\label{eq:BetaEstimate}
\beta=\frac{n_J}{n_Q^2}\,\alpha^{-1}\,\frac{9(g-1)}{5}\,.
\end{equation}
This nicely shows that the slow-rotation approximation constrains 
the given combination of angular momentum, charge, and gyromagnetic 
factor. In particular, any gyromagnetic factor up to $g=6$ can be 
so obtained, given that the charge is sufficiently large. If we set 
$g=2$ and $n_J=1$ (corresponding to the electron's values), we get 
\begin{equation}
\label{eq:BetaEstimateElectron}
n_Q\gg \sqrt{n_J (g-1)247}\approx 16\,.
\end{equation}
This means that indeed we cannot cover the electrons values with the 
present model while keeping the slow-rotation approximation, though 
this model seems to be able to accommodate values of $g$ up to six 
if the charge is sufficiently high. However, we did not check whether 
the assumption that the matter of the shell provided the stabilising 
stresses is in any way violating general conditions to be imposed on 
any energy-momentum tensor. This we shall do next.   

\subsubsection{Restrictions by energy dominance}
\label{sec:RestEnDom}
Energy dominance essentially requires the velocity of sound in the 
stress-supporting material to be superluminal. It is conceivable 
that for certain values of the physical quantities $(m,Q,g,J)$ the 
stresses would become unphysically high. To check that, at least for 
the condition of energy dominance, we first note from 
(\ref{eq:FullEM-Tensor-c}) and (\ref{eq:ElMass}) that the stress 
part of the energy-momentum tensor can be written in the form 
\begin{equation}
\label{eq:T-sigma}   
\mathbf{T}_\sigma=-\,\frac{1}{2}\frac{m_e}{4\pi R^2}c^2\delta(r-R)\mathbf{P}\,.
\end{equation}
Hence the ratio between the stress within the shell (in any direction 
given by the unit spacelike vector $n$ tangent to the shell, so that 
$n\cdot\mathbf{P}\cdot n=1$) and its energy density, as measured by a 
locally co-rotating observer, is given by 
\begin{equation}
\label{eq:RatioStresses}   
\left\vert\frac{n\cdot\mathbf{T}\cdot n}{u\cdot\mathbf{T}\cdot u}\right\vert
=\frac{m_e}{2m_0}
=\frac{3(g-1)}{6-g}\,,
\end{equation}
where we used (\ref{eq:FunctParaObs-a}) and (\ref{eq:FunctParaObs-b})
in the last step. The condition of energy dominance now requires this 
quantity to be bounded above by 1, so that  
\begin{equation}
\label{eq:EnergyDominance}   
\frac{3(g-1)}{6-g}\leq 1\Longleftrightarrow g\leq\frac{9}{4}\,.
\end{equation}
Interestingly this depends on $g$ only. Hence we get, after all, an upper 
bound for $g$, though from the condition of energy dominance, i.e. 
a subluminal speed of sound in the shell material, and not from the 
condition of a subluminal rotational speed.  

\subsubsection{The size of the electron}
\label{eq:ElectronSize}
What is the size of the electron? According to (\ref{eq:FunctParaObs-c}), 
its radius comes out to be  
\begin{equation}
\label{eq:ElectronRadius-1}
R=\frac{1}{4\pi\varepsilon_0 c^2}\,\frac{e^2}{2m}\,\frac{5}{3}\,,
\end{equation}
where we set $Q=-e$ and $g=2$. On the other hand, in Quantum Mechanics, 
the Compton wavelength of the electron is 
\begin{equation}
\label{eq:ElectronComptonW-length}
\lambda=\frac{2\pi\hbar}{mc}\,,
\end{equation}
so that their quotient is just 
\begin{equation}
\label{eq:QuotComptonW-lengthRadius}
\frac{R}{\lambda}=\frac{5}{6}\frac{\alpha}{2\pi}\approx 2\cdot 10^{-3}\,.
\end{equation}
This might first look as if the classical electron is really small,
at least compared to its Compton wavelength. However, in absolute 
terms we have ($\mathrm{fm}$ stands for the length scale ``Fermi'')
\begin{equation}
\label{eq:ElectronRadius-2}
R\approx 2\cdot 10^{-15}\mathrm{m}=2\,\mathrm{fm}\,,
\end{equation}
which is very large compared to the scale of $10^{-3}\,\mathrm{fm}$ 
at which modern high-energy experiments have probed the electron's 
structure, so far without any indication for substructures. 
At that scale the model discussed here is certainly not capable of 
producing any reasonable values for the electron parameters, since the 
electrostatic mass (and hence the total mass, if we assume the weak 
energy-condition, $m_0>0$, for the shell matter) comes out much too 
large and the angular momentum much too small (assuming $\beta<1$).

One might ask whether the inclusion of gravity will substantially 
change the situation. For example, one would expect the gravitational 
binding to reduce the electrostatic self-energy.  An obvious and 
answerable questions is whether the electron could be a Black Hole? 
What is particularly intriguing about spinning and charged Black Holes 
in General Relativity is that their gyromagnetic factor is $g=2$, 
always and exactly!\footnote{It is known that $g=2$ is already a 
preferred value in special-relativistic 
electrodynamics~\cite{Bargmann.Michel.Telegdi:1959}, a fact on which 
modern precision measurements of $g-2$ rest. See \cite{Pfister.King:2003} 
and \cite{Garfinkle.Traschen:1990} for instructive discussions as to 
what makes $g=2$ also a special value in General Relativity.} For a 
mass $M$ of about $10^{-30}\,\mathrm{kg}$ to be a Black Hole it must be 
confined to a region smaller than the Schwarzschild radius 
$R_s=2GM/c^2\approx 10^{-57}\,\mathrm{m}$, which is almost 40 orders 
of magnitude below the scale to which the electron structure has been 
probed and found featureless. Hence, leaving alone Quantum Theory, it 
is certainly a vast speculation to presumes the electron to be a Black 
Hole. But would it also be inconsistent from the point of view of General 
Relativity? The Kerr-Newman family of solutions for the Einstein-Maxwell
equations allow any parameter values for mass (except that it must be 
positive), charge, and angular momentum. As already stated, $g=2$ 
automatically. Hence there is also a solution whose parameter values 
are those of the electron. However, only for certain restricted ranges 
of parameter values do these solutions represent Black Holes, that is, 
possess event horizons that cover the interior singularity; otherwise 
they contain naked singularities. 

More precisely, one measures the mass $M$, angular momentum per unit 
mass $A$, and charge $Q$ of a Kerr-Newman solution in geometric units, 
so that each of these quantities acquires the dimension of length. 
If we denote these quantities in geometric units by the corresponding 
lower case letters, $m$, $a$, and $q$ respectively, we have 
\begin{subequations}
\label{eq:GeometricUnits}
\begin{alignat}{2}
\label{eq:GeometricUnits-Mass}
&m&&\,=\,M\,\frac{G}{c^2}\,,\\
\label{eq:GeometricUnits-AngMom}
&a&&\,=\,\frac{A}{c}\,,\\
\label{eq:GeometricUnits-Charge}
&q&&\,=\,Q\,\sqrt{\frac{\mu_0}{4\pi}\,\frac{G}{c^2}}\,.
\end{alignat}
\end{subequations}
The necessary and sufficient condition for an event horizon to 
exist is now given by 
\begin{equation}
\label{eq:EventHorCond}
\left(\frac{a}{m}\right)^2+\left(\frac{q}{m}\right)^2\leq 1\,. 
\end{equation}
The relevant quantities to look at are therefore the dimensionless 
ratios\footnote{We write $P[X]$ to denote the number that 
gives the physical quantity $P$ in units of $X$.}
\begin{subequations}
\label{eq:BH-Ratios}
\begin{alignat}{5}
\label{eq:BH-Ratios-a}
&\frac{a}{m}&&\quad=\quad &&\frac{A}{M}&&\cdot\frac{c}{G}
&&\quad\approx\quad\frac{A[\mathrm{m^2\cdot s^{-1}}]}%
{M[\mathrm{kg}]}\cdot 5.5\cdot 10^{18}\,,\\
\label{eq:BH-Ratios-b}
&\frac{q}{m}&&\quad=\quad &&\frac{Q}{M}&&\cdot\sqrt{\frac{\mu_0}{4\pi}\,\frac{c^2}{G}}
&&\quad \approx\quad \frac{Q[\mathrm{C}]}{M[\mathrm{kg}]}\cdot 10^{10}\,.
\end{alignat}
\end{subequations}
Now, if we insert the parameter values for the electron\footnote{We have 
$A=S/M$ with $S=\frac{1}{2}\hbar$ (modulus of electron spin) and 
use the approximate values $\hbar[\mathrm{J\cdot s}]\approx 10^{-34}$, 
$M[\mathrm{kg}]=10^{-30}$, and $Q[\mathrm{C}]=1.6\cdot 10^{-19}$.}
(we take for $Q$ the modulus $e$ of the electron charge) we arrive 
at the preposterous values

\begin{subequations}
\label{eq:ElectronBH-Ratios}
\begin{alignat}{3}
\label{eq:NumericElectronBH-Ratios-a}
&\frac{a}{m}\Big\vert_{\mathrm{electron}}
&&\quad\approx\quad\Bigl(5\cdot 10^{25}\Bigr)\Bigl(5.5\cdot 10^{18}\Bigr)
&&\quad\approx\quad 2.5\cdot 10^{44}\,,\\
\label{eq:NumericElectronBH-Ratios-b}
&\frac{q}{m}\Big\vert_{\mathrm{electron}}
&&\quad\approx\quad\Bigl(1.6\cdot 10^{11}\Bigr)\cdot 10^{10}
&&\quad\approx\quad 1.6\cdot 10^{21}\,,
\end{alignat}
\end{subequations}
so that we are indeed very far from a Black Hole. Classically 
one would reject the solution for the reason of having a naked 
singularity. But note that this does not exclude the possibility that 
this exterior solution is valid up to some finite radius, and is then 
continued by another solution that takes into account matter sources 
other than just the electromagnetic field.\footnote{Even in mesoscopic 
situations $a<m$ means a very small angular momentum indeed. Recall 
that in Newtonian approximation the angular momentum of a homogeneous 
massive ball of radius $R$ is $2MR^2\omega/5$, so that $a/m\leq 1$ 
translates to the following inequality for the spin period 
$T=2\pi/\omega$: 
\begin{equation}
\label{eq:MacrBodiesAM-Ratio}
T\geq\frac{4\pi}{5}\,\frac{R}{c}\,
\frac{R}{m}\approx \frac{R^2[\mathrm{m}]}{M[\mathrm{kg}]}\cdot 10^{19}\,
\mathrm{sec}\,,
\end{equation}
which for a ball of radius one meter and mass $10^{3}$ kilogrammes 
sets an upper bound for $T$ of $3\cdot 10^8\,\mathrm{years}$! In fact, 
(\ref{eq:MacrBodiesAM-Ratio}) is violated by all planets in 
our solar system.}

\section{Summary}
\label{sec:Summary}
Understanding the generation of new ideas and the mechanisms that 
led to their acceptance is a common central concern of historians of
science, philosophers of science, and the working scientists themselves.
The latter might even foster the hope that important lessons can be 
learnt for the future. In any case, it seems to me that from all 
perspectives it is equally natural to ask whether a specific argument 
is actually true or just put forward for persuasive reasons. 

Within the history of Quantum Mechanics the history of spin is, in 
my opinion, of particular interest, since it marks the first instance 
where a genuine quantum degree of freedom without a classically 
corresponding one were postulated to exist. If this were the general 
situation, our understanding of a quantum theory as the quantisation 
of a classical theory cannot be fundamentally correct.\footnote{I take 
this to be an important and very fundamental point. Perhaps with the 
exception of Axiomatic Local Quantum Field Theory, any quantum theory 
is in some sense the quantisation of a classical theory. Modern 
mathematical theories of `quantisation' understand that term as 
`deformation' (in a precise mathematical sense) of the algebra of 
observables over \emph{classical} phase space; cf.~\cite{Waldmann:PGuD}.} 
On the other hand, modern theories of quantisation can explain the 
quantum theory of a spinning particle as the result of a quantisation 
applied to some classical theory, in which the notion of spin is already 
present.\footnote{Namely in the sense that it has a corresponding 
classical state space given by a two-sphere, which is a symplectic 
manifold. However, this state space is not the phase space (i.e. 
cotangent bundle) over some space of classical configurations, so that 
one might feel hesitant to call it a classical \emph{degree of freedom}.}
Hence, from a modern perspective, it is simply not true that spin has no 
classical counterpart. That verdict (that is has no classical counterpart),
which is still often heard and/or read\footnote{Even in critical historical 
accounts, e.g.: ``Indeed, there were unexpected results from quantum theory 
such as the fact that the electron has a fourth degree of freedom, namely, 
a spin which has no counterpart in a classical theory'' (\cite{Miller:1973},
p.\,319).}, is based on a narrow concept of `classical system', which 
has been overcome in modern formulations, as was already 
mentioned in footnote\,\ref{foot:ClassicalSpin} to which I refer at 
this point. From that point of view, spin is no less natural in classical 
physics than in Quantum Theory, which has now become the standard attitude
in good textbooks on analytical mechanics, 
e.g. \cite{Souriau:SoDS}\cite{Guillemin.Sternberg:1990} as well as in 
attempts to formulate theories of quantisation~\cite{Waldmann:PGuD}\cite{Woodhouse:GQ}.
      
In the present contribution I concentrated on another aspect, namely 
whether it is actually true that classical models for the electron 
(as they were already, or could have been, established around 1925) 
are not capable to account for the actual values of the four electron 
parameters: mass, charge, angular momentum, and the gyromagnetic factor. 
This criticism was put forward from the very beginning (Lorentz) and 
was often repeated thereafter. It turns out that this argument is not
as clear cut as usually implied. In particular, $g=2$ is by no means 
incompatible with classical physics. Unfortunately, explicit 
calculations seem to have been carried out only in a simplifying 
slow-rotation approximation, in which the Poincar\'e stresses may be 
taken uniform over the charged shell. In the regime of validity of
this approximation $g=2$ is attainable, but not for small charges. 
I do not think it is known whether and, if so, how an exact 
treatment improves on the situation. In that sense, the answer to the 
question posed above is not known. An exact treatment would have to 
account for the centrifugal forces that act on the rotating shell in 
a latitude dependent way. As a result, 
the Poincar\'e stresses cannot retain the simple (constant) form as in 
(\ref{eq:FullEM-Tensor-c}) but must now also be latitude dependent.
In particular, they must be equal in sign but larger in magnitude than 
given in (\ref{eq:SurfaceStress}) since now they need in addition to 
balance the outward pushing centrifugal forces. On one hand, this 
suggests that their effect is a still further reduction of angular 
momentum for fixed magnetic moment, resulting in still larger values 
for $g$. On the other hand, fast rotational velocities result in an 
increase of the inertial mass according to (\ref{eq:MassVelDep})
and hence an increase of angular momentum, though by the same token 
also an increase in the centrifugal force and hence an increase in stress.
How the account of these different effects finally turns out to be is
unclear (to me) without a detailed calculation.\footnote{In that respect 
the corresponding statements made in Sect.\,4.7.1 of 
\cite{Schulte:SpinDiss1999} seem to me premature.} It would be of 
interest to return to this issue in the future.

\addcontentsline{toc}{section}{References}
%\bibliographystyle{plain}
%\bibliography{RELATIVITY,QM,HIST-PHIL-SCI,MATH}

\end{document}